\documentclass{edm_article}
\usepackage[dvipsnames]{xcolor}
\usepackage{url}

\begin{document}

\title{
\textcolor{blue}{\normalsize Appears in the Proceedings of Seventeenth International Conference on Educational Data Mining (EDM) Workshop: Leveraging LLMs for Next Generation Educational Technologies, July 2024.} \\[1em]

How Do Students Interact with an LLM-powered Virtual Teaching Assistant in Different Educational Settings?}

\numberofauthors{2}
\author{
\alignauthor
Pratyusha Maiti\\
       \affaddr{Georgia Institute of Technology}\\
       \affaddr{Atlanta, Georgia, USA}\\
       \email{pmaiti6@gatech.edu}
\alignauthor
Ashok K. Goel\\
       \affaddr{Georgia Institute of Technology}\\
       \affaddr{Atlanta, Georgia, USA}\\
       \email{ashok.goel@cc.gatech.edu}
}     

\maketitle

\begin{abstract}
Jill Watson, a virtual teaching assistant powered by LLMs, answers student questions and engages them in extended conversations on courseware provided by the instructors. In this paper, we analyze student interactions with Jill across multiple courses and colleges, focusing on the types and complexity of student questions based on Bloom's Revised Taxonomy and tool usage patterns. We find that, by supporting a wide range of cognitive demands, Jill encourages students to engage in sophisticated, higher-order cognitive questions. However, the frequency of usage varies significantly across deployments, and the types of questions asked depend on course-specific contexts. These findings pave the way for future work on AI-driven educational tools tailored to individual learning styles and course structure, potentially enhancing both the teaching and learning experience in classrooms.
\end{abstract}

\keywords{Conversational AI Agents, Question Answering, Virtual Teaching Assistant} 

\section{Introduction}
The integration of Generative AI tools into classroom settings marks a significant evolution in educational technologies, shifting towards more interactive and personalized learning experiences. As these technologies become increasingly prevalent, understanding the dynamics of real interactions between students and AI becomes crucial. This understanding not only helps in assessing the effectiveness of AI tools in enhancing learning outcomes but also in refining their design to better address student needs and behaviors.

Jill Watson is a conversational vitual teaching assistant that answers students questions and engages them in extended conversations based on teacher-prescribed courseware using LLMs and retrieval augmented generation in the backend\cite{taneja2024jill}.
Jill Watson has been equipped with OpenAI's GPT-3.5 Turbo model, accessed via the OpenAI API, and coupled with several other technologies to facilitate more nuanced, context-aware, and safe interactions with students.
Jill has been deployed in both online and offline classrooms\cite{kakar2024jill} across different educational institutes and courses.
This paper examines student interactions with Jill Watson, to understand how AI-based educational tools may engage students in meaningful and deeper learning experiences. 



The paper has two main contributions: 
(i) We report trends in usage of an LLM-powered conversational question-answering (QA) agent  observed through organic interactions
of students with the tool deployed in classroom settings. This analysis provides insights into how students interact with the tool naturally, without guided prompts or interventions.
(ii) Our findings reveal that by supporting a diverse range of student questions across varying levels of cognitive complexity, the QA agent facilitates and encourages students to pose more complex, higher-order questions.
Such interactions not only foster deeper cognitive involvement but also potentially boost student motivation by engaging them in more challenging and intellectually stimulating discussions.

In subsequent sections, we provide theoretical motivation for our evaluation and describe the datasets, tools and analysis employed to study the impact of Jill Watson on frequency of usage and complexity of interactions.
Section 2 explores previous work on student usage of AI tools in classrooms and discusses our work in this context.
Section 3 details the various sources of data and tools for analyzing student interaction with the tool as well as the evaluation metrics focusing on the trends of tool usage and the cognitive dimensions of student queries.
In section 4, we present our findings and compare usage trends across two different educational settings.
We conclude with a summary of key takeaways, limitations and future work on studying the impact of AI tools in classrooms.

\section{Related Work}
Since the introduction of ChatGPT by OpenAI\footnote{\url{https://openai.com/blog/ChatGPT/}} in November 2022, there has been a rising interest in exploring its applications within educational settings. The capability of ChatGPT to mimic human-like conversations and generate coherent, detailed responses\cite{zhai} has been particularly noted for its potential in educational environments. Previous research\cite{chung} categorizes potential education functions for ChatGPT to work as a virtual teaching assistant into learning support — answering questions, summarizing, and facilitating collaboration — and assessment tools — concept checking, exam preparation, drafting, and feedback provision. Empirical studies such as those conducted by Gilson et al.\cite{gilson} observed that responses generated by ChatGPT are structured in a way that they could lead to more in-depth questioning and stimulate students to use their knowledge and reasoning abilities. Conversely, Rudolph et al.\cite{rudolph} warned against relying on ChatGPT as a substitute for critical thinking and originality. Despite several challenges, studies identify the potential of using ChatGPT and similar generative AI tools in classrooms as learning aid in form of teaching assistants or personalized tutors\cite{adiguzel}.

Previously, Goel and Polepeddi introduced a virtual teaching assistant, Jill Watson\cite{goel_2018, goel} built on top of IBM’s Watson platform. Jill answered students’ questions on course logistics in online discussion forums and was trained on historic human TA-student interactions. 

The current version of Jill Watson is designed to engage students in extended conversations on verified courseware for a classroom. Taneja et al.\cite{taneja2024jill} discuss the foundational architecture of Jill Watson, highlighting the design decisions that made Jill's responses contextually relevant, safe and robust to toxic prompting. Kakar et al.\cite{kakar2024jill} introduced modifications to this architecture to tailor Jill for classroom use. Our study focuses on student interactions with Jill deployed in real classrooms along two dimensions - usage patterns and the nature of questions posed by students, thereby enhancing our understanding of Jill's educational impact and effectiveness.

The existing literature predominantly relies on surveys to examine student perceptions of such tools. These studies provide valuable insights into students' attitudes and self-reported experiences but may not fully capture the nuances of real-time, interactive dynamics. For instance, previous research often focuses on students' perceived benefits and drawbacks without observing how students actually use and react to these tools in live scenarios\cite{chan}. In this paper, we report trends in usage observed through organic interactions of students with the AI tool in classrooms.

The application of Bloom’s Taxonomy\cite{bloom1956taxonomy} in educational settings has been extensively explored, mainly for designing educational objectives and assessment tasks. The original taxonomy consists of six levels — Knowledge, Comprehension, Application, Analysis, Evaluation, and Synthesis. Each level represents a step in the cognitive process, from basic recall of facts to higher-order thinking skills such as critical analysis and creative problem-solving. Krathwohl (2002)\cite{krathwohl} provided a foundational revision of Bloom’s original cognitive taxonomy by emphasizing the dynamic nature of learning objectives and categorized them into cognitive processes that are crucial for structuring educational content and assessments. We have adapted this framework to analyze the cognitive complexity of student questions. Cognitive complexity refers to the level of mental processes required to perform a task or understand a concept, ranging from basic recall and comprehension to higher-order thinking skills such as analysis, synthesis, and evaluation. Questions that demand higher-order thinking skills reveal a deeper level of understanding and engagement with the educational content\cite{forehand}.

\section{Methodology}

In this paper, we will examine student interactions from real classroom deployments with Jill Watson.
Our goal is to understand how desirable characteristics of LLMs enhance tool usage and whether inherent adaptability of LLMs has the capability to engage students in critical thinking.

To achieve this, we have divided our work into three focus areas: (1) investigate usage frequencies, (2) explore tool utility by examining which document students ask most questions from and examining the distribution of these questions over time throughout the semester, and (3) explore the complexity of questions that students ask from the tool under different educational settings. In our context, an education setting refers to a distinct classroom that Jill is deployed in, marked by the course structure and contents, and the studdent body taking the course.  
In the subsequent sections, we will expand on each of these focus areas from the point of view of theory, data, and tools used for analysis.
%


\subsection{Agent Design}
%
To examine the effect of LLM integration into Jill Watson on frequency of usage, we will explore student interactions with Jill Watson both pre and post LLM integration. 
Hence, we will briefly describe the architecture for both these versions of the tool to highlight how LLM integration significantly enhanced tool functionality and how it reflects on adoption. 
Table \ref{tab:deploy} shows which version of Jill has been deployed in the academic semesters under study.
To provide context for our findings later in the paper, we will also describe the courses where Jill has been deployed.

\begin{figure}
\Description{Architecture for old Jill}
\centering
\includegraphics[width=2.75in]{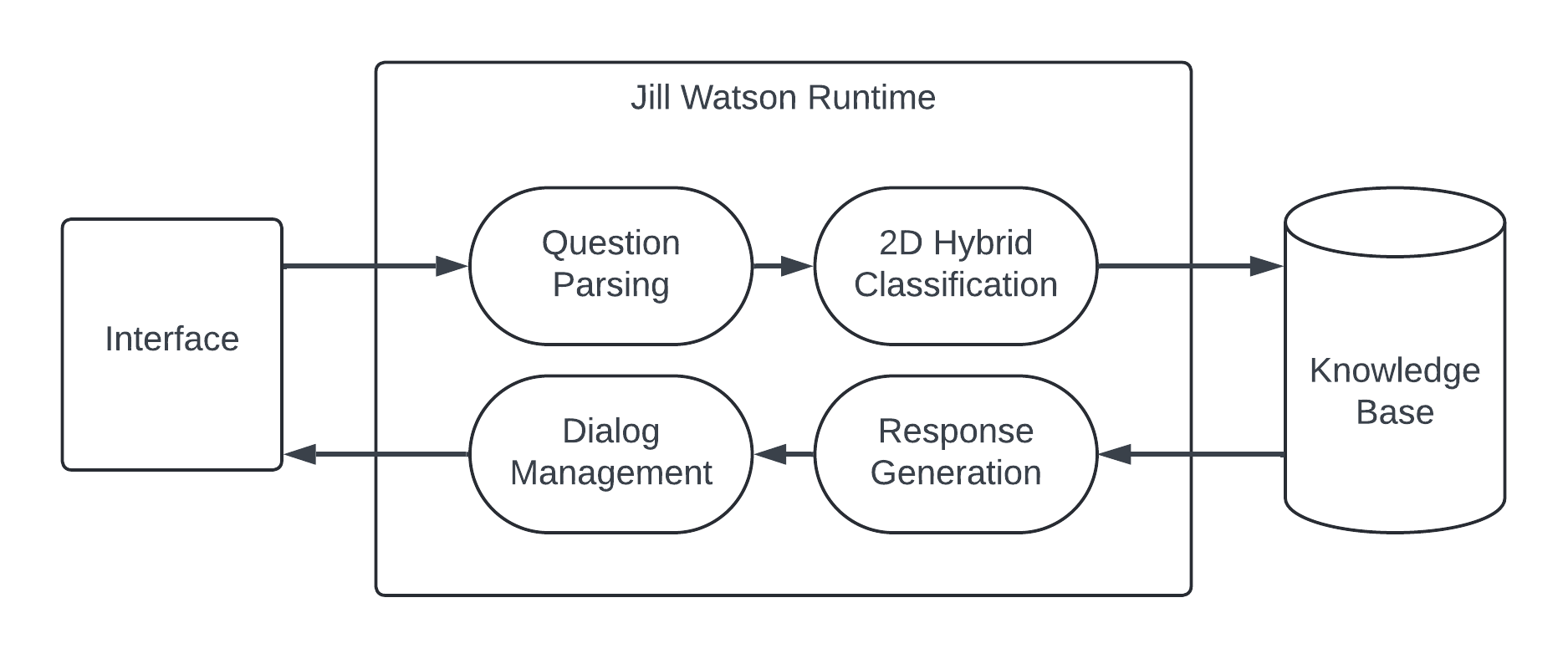}
\caption{Architecture of Jill Watson prior to LLM integration. Adapted from Goel 2022.}
\label{fig:legacy-jill-arch}
\end{figure}

\subsubsection{Legacy Jill Watson}
Pre-LLM version of Jill\cite{askjill} employed a two-dimensional hybrid classification process as shown in Figure \ref{fig:legacy-jill-arch}. A pre-trained NLP model functioning as an intent classifier classifies student questions into predefined question categories. A rule-based semantic processor utilizes the classified intent to select an appropriate response template for generating responses. This iteration of Jill Watson was extensively deployed in classes at Georgia Institute of Technology. For our study, we analyze interaction and testing data from deployment of this version in one particular classroom - the Knowledge-Based AI (KBAI) course at Georgia Tech to understand how the use of Jill Watson has evolved with the introduction of more advanced technologies and LLMs.

\begin{figure}
\Description{Architecture for new Jill}
\centering
\includegraphics[width = 3in]{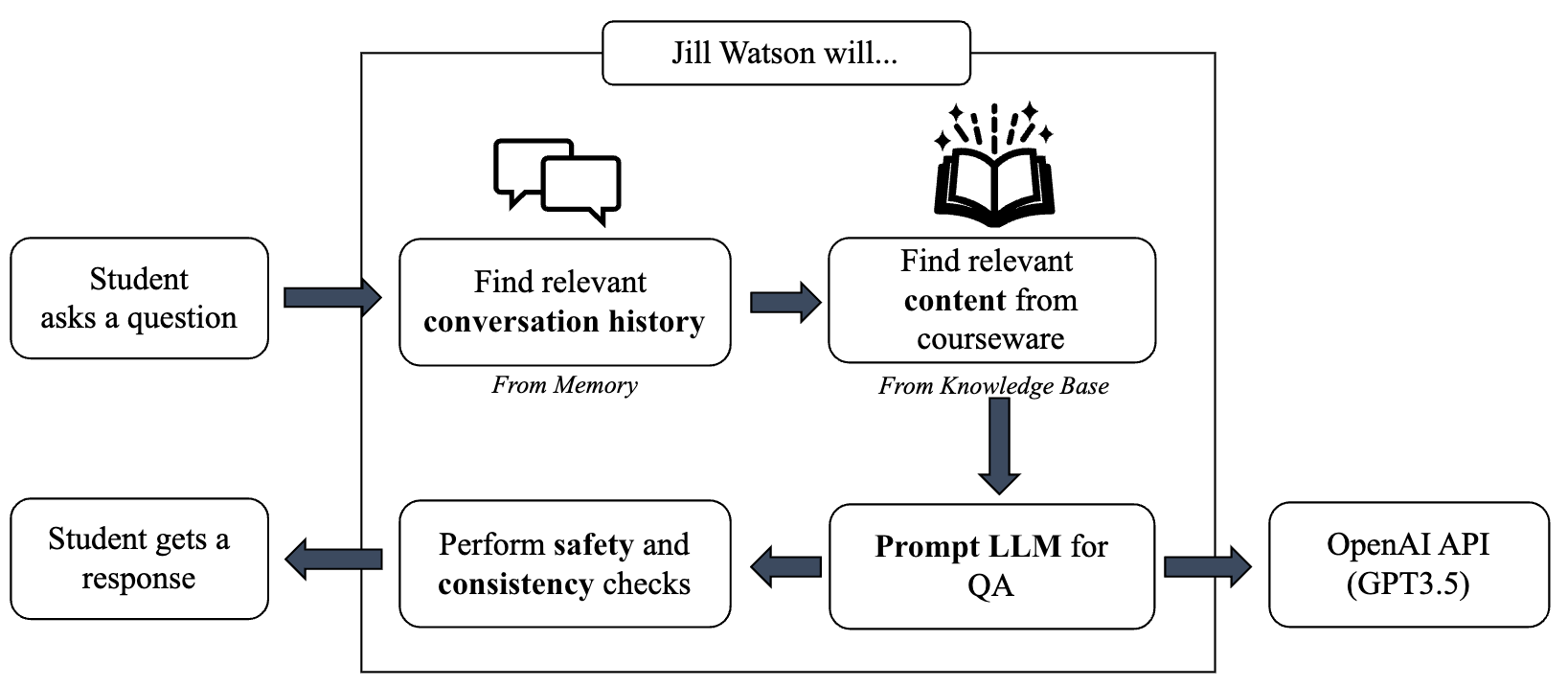}
\caption{Architecture of Jill Watson post LLM integration. Adapted from Taneja 2024.}
\label{fig:jill-arch}
\end{figure}

\subsubsection{LLM-powered Jill Watson}
Post LLM integration, Jill Watson, described by Taneja et al.\cite{taneja2024jill}, features a modular, Retrieval Augmented Generation (RAG)-based pipeline for question-answering. Figure \ref{fig:jill-arch} shows a simplified version of the architecture that provides an intuitive understanding of the LLM-powered QA process. Jill retrieves conversation history for a student from memory, relevant information from the knowledge base (which is pre-configured by processing verified courseware received from instructors), and constructs a prompt for OpenAI API to generate a response restricted to the given context. The response is validated before being sent to the student. By restricting the context to relevant information sourced exclusively from verified course materials and combining moderation modules to ensure the safety and accuracy of responses, Jill goes beyond ChatGPT in response relevancy, accuracy and safety, which are further elaborated in \cite{taneja2024jill}. Further design modifications to tailor Jill for classroom integration are provided by Kakar et al.\cite{kakar2024jill}. These adjustments facilitate Jill Watson's deployment across diverse educational environments, providing rich insights into student interactions. 

\subsubsection{Deployments}
Table \ref{tab:deployments} lists the courses where LLM-powered Jill has been deployed in Spring 2024, the institutes where these courses are taught, and the courseware to answer student questions, and the main course deliverables. These details serve to provide context for the variability within each deployment.
From Fall 2021 through Fall 2022, the legacy version of Jill Watson was deployed in the Knowledge-Based AI (KBAI) course at Georgia Tech. Starting in Fall 2023, the updated version of Jill Watson has been deployed not only in the KBAI course but also in the Introduction to Cognitive Science (CogSci) course at Georgia Tech, and expanded to the English: Composition and Rhetoric course at Wiregrass Georgia Technical College.

The KBAI and Cognitive Science courses are part of Georgia Tech's Online Master of Science in Computer Science program{\footnote{\url{https://omscs.gatech.edu/}}}, designed primarily for part-time graduate students who typically maintain full-time employment. These courses aim to provide advanced knowledge and skills in AI and Cognitive Science through a flexible online format.
In contrast, Wiregrass College, a two-year community college within the Technical College System of Georgia (TCSG) recognized for its workforce development programs, integrates Jill Watson into its undergraduate English course. The course teaches various modes of writing and includes a review of standard grammatical and stylistic usage in proofreading and editing.

In each of these courses, Jill Watson is deployed as a private conversational assistant within the Learning Management System (LMS) using Learning Tools Interoperability (LTI)\footnote{\url{https://www.1edtech.org/standards/lti}}. Jill can be accessed through the course's Canvas\footnote{\url{https://www.instructure.com/canvas/}} or Blackboard\footnote{\url{https://www.blackboard.com/}} page.
This integration ensures that students can access assistance directly within their course platform, encouraging a seamless and supportive learning experience. Once a question is asked on the platform, a student typically receives a structured response within a few seconds. Much of the delay is caused due to multiple calls to OpenAI’s API.

\begin{table}[]
\centering
\begin{tabular}{|l|l|l|}
\hline
\textbf{Semesters}                                                 & \textbf{Deployed Version} & \textbf{Supported Material}                                                      \\ \hline
\begin{tabular}[c]{@{}l@{}}Fall 2021 - \\ Fall 2022\end{tabular}   & pre-LLM               & Syllabi                                                                            \\ \hline
\begin{tabular}[c]{@{}l@{}}Fall 2023 - \\ Spring 2024\end{tabular} & post-LLM              & \begin{tabular}[c]{@{}l@{}}Syllabi, textbooks\end{tabular} \\ \hline
\end{tabular}
\caption{Jill Watson deployments in KBAI class used for this study. The number of enrolled students across the semesters is consistently about 200.}
\label{tab:deploy}
\end{table}

\subsection{Data Collection}

For this study, we use data from three different sources: (i) student interaction data from deployments of the LLM-powered Jill Watson in Spring 2024, and (ii) student interaction data from deployments of the pre-LLM version of Jill Watson in the KBAI class from Fall 2021 to Fall 2022
and (iii) a set of synthetically generated context-question-response tuples based on various courseware for Spring 2024.

\subsubsection{Student Interaction with Jill Watson}
We collected and stored exhaustive student interaction data for each deployment of Jill Watson in various courses in a persistent memory on the hosting server, following established protocols. This dataset includes de-anonymized student IDs, timestamps, questions, responses, and feedback, along with inputs and outputs for each component of the question-and-answer (QA) pipeline. This structured data collection allows for a robust analysis of usage patterns.

\begin{table}[]
\centering
\begin{tabular}{|l|l|l|l|}
\hline
\textbf{Inst} & \textbf{Course} & \textbf{\begin{tabular}[c]{@{}l@{}}Supported \\ courseware\end{tabular}}     & \textbf{\begin{tabular}[c]{@{}l@{}}Course\\ Structure\end{tabular}}                            \\ \hline
GT            & KBAI            & \begin{tabular}[c]{@{}l@{}}Syllabus,\\ Textbook\end{tabular}                      & \begin{tabular}[c]{@{}l@{}}Coding assignments,\\ journal writing, exams\end{tabular}           \\ \hline
GT            & CogSci          & \begin{tabular}[c]{@{}l@{}}Syllabus,\\ Textbook,\\  Video \\Transcripts\end{tabular} & \begin{tabular}[c]{@{}l@{}}Weekly quizzes,\\ writing assignments,\\ final project\end{tabular} \\ \hline
TCSG            & English             & \begin{tabular}[c]{@{}l@{}}Syllabus,\\ Textbook,\\ MLA Guide\end{tabular}         & Writing assignments                                                                            \\ \hline
\end{tabular}
\caption{Jill Watson is deployed across two institutes Georgia Tech (GT) and Technical College System of Georgia (TCSG).}
\label{tab:deployments}
\end{table}


\subsubsection{Student Interaction with Legacy Jill}
We have sourced the number of questions asked of legacy Jill Watson in the KBAI course spanning from Fall 2021 to Fall 2022. We will use these numbers to compare the frequency of usage with the current version of Jill. 


\subsection{Pre-deployment Testing of Jill Watson}
We generate synthetic context-question-answer (CQA) tuples using Google's PaLM API for supported courseware, ranging a variety of documents like syllabi, textbook, instructor handouts and guides across five different courses. These question sets are used to test Jill's accuracy for courseware for a given class prior to deployments. We prompt the API to create a diverse set of questions that cover a range of cognitive levels and types (e.g., factual, conceptual, procedural) as well as different question types (e.g., is, which, when, where, why, how). To ensure these CQAs are representative of the questions Jill Watson is expected to address, we incorporate guidelines that require the API to generate questions needing multiple pieces of information and to reference specific values within the course materials. This approach helps simulate the complexity and diversity of real student inquiries. These synthetic CQAs are used as a benchmark to compare the expected versus actual usage of Jill Watson in classroom settings.

\subsection{Classifying Question Complexity with Fine-Tuned BERT}
To understand the type and complexity of questions asked in classrooms, we use cognitive levels defined by Bloom's Revised Taxonomy\cite{krathwohl}. We compiled a dataset of 3,600 labeled questions derived from various publicly accessible sources\cite{data2, data1}. This dataset contains questions across various disciplines and labels them into one of the six cognitive levels defined in the taxonomy - Knowledge, Comprehension, Application, Analysis, Synthesis, and Evaluation, which range from basic recall of facts to generating new ideas. 

First, we investigated zero-shot classification techniques using GPT-3.5, GPT-4, and BERT for categorizing student queries. Given the nuanced nature of this classification, zero-shot methods proved ineffective, with accuracies lower than 50\%, prompting us to explore fine-tuning strategies on a curated dataset to improve performance.

For fine-tuning, we divided the dataset with 3600 questions into training and testing subsets with a 75:25 split. Standard preprocessing methods were applied to each question to ensure consistency across the data. These methods included tokenization, stopwords removal, and text normalization, which were essential for minimizing variability and enhancing the efficiency of model training.

We opted for a fine-tuning strategy on the \textit{bert-base-uncased} BERT model to classify questions based on cognitive complexity. The model underwent fine-tuning over five epochs, using a batch size of 32 and a learning rate of 2e-5, parameters selected from initial tests that indicated optimal results. We settled on using the fine-tuned model that achieved an accuracy of 91.9\%.

\section{Results and Discussion}
In this section, we analyze usage patterns, interaction frequencies, the complexity of questions, and the qualitative aspects of Jill Watson's responses, to provide a comprehensive understanding of how LLMs enhance the educational experience.
Each subsection addresses specific aspects of the tool's performance and interaction with students across different educational settings. 

\begin{table}[]
\centering
\begin{tabular}{|l|l|l|l|}
\hline
\textbf{Criteria}                                                    & \textbf{\begin{tabular}[c]{@{}l@{}}Sp24\\ KBAI\end{tabular}} & \textbf{\begin{tabular}[c]{@{}l@{}}Sp24\\ CogSci\end{tabular}} & \textbf{\begin{tabular}[c]{@{}l@{}}Sp24\\ English\end{tabular}} \\ \hline
Total enrolled students                                              & 223                                                          & 58                                                             & 60                                                              \\ \hline
\begin{tabular}[c]{@{}l@{}}Used Jill Watson \\ atleast once\end{tabular} & 34.0\%                                                        & 53.5\%                                                          & 15.0\%                                                           \\ \hline
\begin{tabular}[c]{@{}l@{}}Used Jill Watson \\ > 5 times\end{tabular}    & 17.5\%                                                        & 27.6\%                                                          & 0                                                               \\ \hline
\begin{tabular}[c]{@{}l@{}}Used Jill Watson \\ > 10 times\end{tabular}   & 11.7\%                                                        & 13.8\%                                                          & 0                                                               \\ \hline
\begin{tabular}[c]{@{}l@{}}Used Jill Watson \\ > 20 times\end{tabular}   & 9.8\%                                                         & 8.6\%                                                           & 0                                                               \\ \hline
\begin{tabular}[c]{@{}l@{}}Total questions asked\end{tabular}   & 1920                                                         & 332                                                           & 32                                                               \\ \hline
\end{tabular}
\caption{Frequency of usage for Jill Watson deployed in classrooms in Spring 2024 (Sp24).}
\label{tab:usage_freq}
\end{table}

\begin{table*}
\caption{Student race/ethnicity across educational settings: Georgia Tech and Wiregrass College(TCSG)}
\centering
\label{tab:dist_race}
\begin{tabular}{|l|l|l|l|l|l|l|l|}
\hline
\textbf{Educational Setting} &
  \textbf{\begin{tabular}[c]{@{}l@{}}American \\ Indian\\ /Alaskan \\ Native\end{tabular}} &
  \textbf{Asian} &
  \textbf{\begin{tabular}[c]{@{}l@{}}Black/\\ African \\ American\end{tabular}} &
  \textbf{\begin{tabular}[c]{@{}l@{}}Hispanic\\ /Latino\end{tabular}} &
  \textbf{\begin{tabular}[c]{@{}l@{}}Two \\ or \\ more\end{tabular}} &
  \textbf{Unk} &
  \textbf{White} \\ \hline
\begin{tabular}[c]{@{}l@{}}Georgia Tech\end{tabular} &
  0 &
  51.25\% &
  3.54\% &
  6.45\% &
  2.71\% &
  2.50\% &
  33.54\% \\ \hline
\begin{tabular}[c]{@{}l@{}}Wiregrass College\end{tabular} &
  1.00\% &
1.00\% &
  40.00\% &
  11.50\% &
  4.5\% &
  0 &
  42.00\% \\ \hline
\end{tabular}
\end{table*}

\subsection{Usage Patterns}

\begin{figure}
\centering
\Description{Student engagement in the KBAI class across semesters showing marked increase after the deployment of LLM-powered Jill Watson.}
\includegraphics[width=3in]{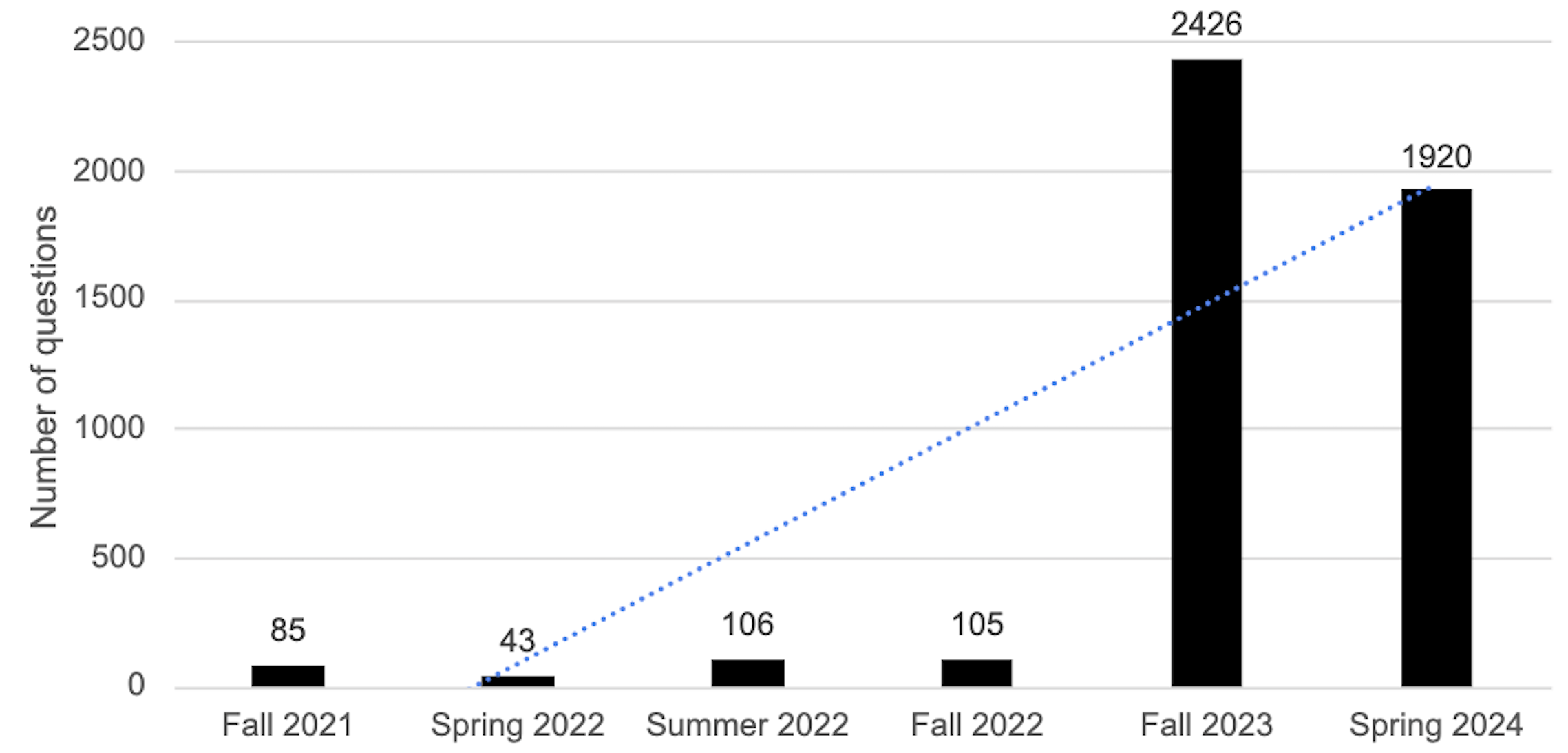}
\caption{Student engagement in the KBAI class across semesters showing marked increase after the deployment of LLM-powered Jill Watson.}
\label{fig:usage_KBAI}
\end{figure}

To explore the breadth of tool usage, we compared the number of questions asked in the KBAI class across six semesters, showing a substantial increase post-LLM integration. The average number of questions rose from 84.75 pre-LLM to 2173.0 post-LLM, as seen in Figure \ref{fig:usage_KBAI}, likely due to the enhanced capabilities of Jill Watson and increased student familiarity with AI tools.

To understand our sample size and set the stage for further analysis of student interaction with Jill, we look at the frequency of usage and number of unique users that used Jill multiple times over the course of the Spring 2024 semester. Table \ref{tab:usage_freq} shows the total number of enrolled students in each class and the percentage of students who used Jill at least once, more than 5, 10 and 20 times. The table also shows the total number of questions asked per course.

We find that Intro to Cognitive Science students exhibited the highest usage, with 53.5\% of the students using the tool at least once, and a notable percentage continuing to use it more frequently (27.6\% used it more than 5 times, and 8.6\% more than 20 times). Similar numbers are observed for the KBAI class. We also note a significant number of questions asked throughout the semester for both these classes. This high level of consistent usage suggests that the tool is particularly effective for both courses. However, while courses at Georgia Tech show consistently high engagement, we observe a significantly lower engagement from students in the English course at Wiregrass College.

%


We hypothesize that several factors such as demographic differences and varying educational motivations between the two student groups may contribute to this phenomenon. Tables \ref{tab:dist_race},  \ref{tab:dist_gender} and \ref{tab:dist_age} show how starkly different the two demography groups are. However, confirming this hypothesis requires further study that is out of the scope of this paper.

\begin{table}[]
\centering
\caption{Student Gender across educational settings: Georgia Tech (GT) and  Wiregrass College(TCSG)}
\label{tab:dist_gender}
\begin{tabular}{|l|l|l|}
\hline
\textbf{Educational Setting}                    & \textbf{Female} & \textbf{Male} \\ \hline
Georgia Tech             & 28.54\%             & 71.46\%           \\ \hline
Wiregrass College & 78.00\%             & 22.00\%            \\ \hline
\end{tabular}

\end{table}

\begin{table}[]
\centering
\caption{Student age across educational settings: Georgia Tech (GT) and  Wiregrass College(TCSG)}
\label{tab:dist_age}
\begin{tabular}{|l|l|l|l|l|l|}
\hline
\textbf{F23}                                                              & \textbf{17-24} & \textbf{25-34} & \textbf{35-44} & \textbf{45-54} & \textbf{55-64} \\ \hline
GT                                                           & 12.50\%             & 64.79\%            & 17.08\%             & 4.16\%             & 1.45\%              \\ \hline
\begin{tabular}[c]{@{}l@{}}TCSG\end{tabular} & 67.50\%            & 23.50\%             & 6.00\%             & 2.00\%              & 1.00\%              \\ \hline
\end{tabular}
\end{table}

\subsection{Interaction Patterns}

Since we have established that Jill has been used extensively in courses, we next look at which category student questions fall into, and how they vary across the semester.
Jill has been equipped to answer questions based on courseware which can be bucketed into two main categories - course content which includes textbooks, guides, handouts and course logistics which include course syllabi and schedule. 

By analyzing the count of questions related to logistics versus content on a weekly basis, we gain a preliminary understanding of the usage patterns. Logistics questions tend to be straightforward, due-diligence queries that help students navigate the course structure. Higher frequency early in the semester typically reflects students' efforts to understand course logistics. Content questions indicate deeper cognitive engagement, as students seek to understand and apply course material. High frequency of content questions suggests heightened cognitive engagement.

\begin{figure*}
\centering
\Description{Time series of the number of questions asked bucketed by relevant courseware for KBAI in Spring 2024}
\includegraphics[width=5.8in]{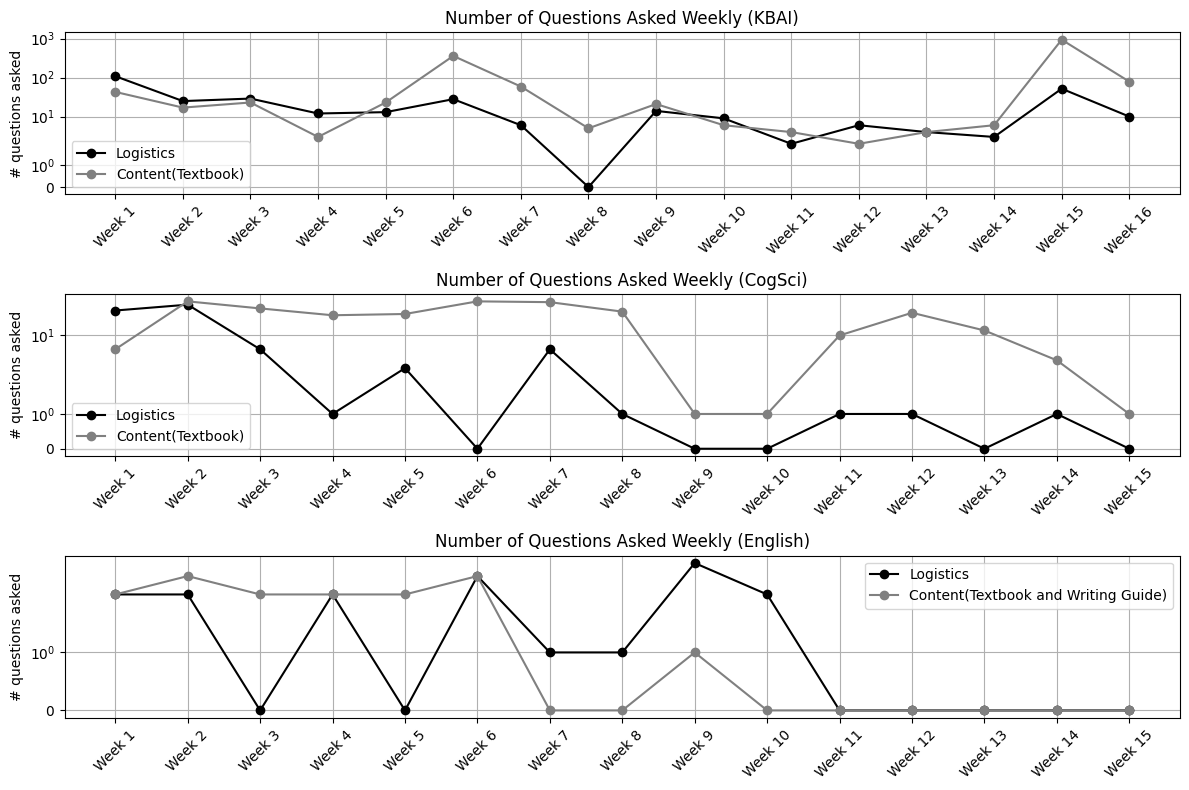}
\caption{Number of questions asked weekly for each course in Spring 2024, bucketed by the document type: Logistics or Content - we see spikes in questions from course content during exam weeks in KBAI (top), relatively consistent engagement with course content in CogSci (middle) and overall lower interaction, with sporadic peaks, in English (bottom).}
\label{fig:time_series_combined}
\end{figure*}

Figure \ref{fig:time_series_combined} shows the number of questions asked weekly for each course in Spring 2024. 
For KBAI, we observe a substantial spike in 'course content' inquiries during week 8 and week 17, which coincide with the examination periods, suggesting that students find Jill useful for looking up information for exam preparation or testing their understanding of concepts.
The significant increase in usage also coincides with the increased complexity of topics covered as the semester advances.

In contrast, the Intro to Cognitive Science course demonstrates significantly higher engagement, particularly in the 'course content' category, with consistent activity throughout the semester. The peak in logistics-related questions during the early weeks may reflect an initial exploration of concepts before stabilizing as the semester progresses. The sustained high volume of content-related questions underscores the tool's effectiveness in supporting complex conceptual understanding and course-specific learning.
The data for the English course shows relatively low overall interaction across all categories, with notable peaks in 'course content' queries during the mid-semester weeks.

Our observations suggest two takeaways. 
First, we notice higher frequency of logistics questions early in the semester which is unsurprising as students navigate the course.
Second, in the KBAI and CogSci classes, there is a clear pattern of high content-related questions, but the pattern differs with course structure.
Courses with evenly distributed deliverables might encourage consistent engagement and tool usage, leading to a more steady learning process. Conversely, courses with concentrated assessments might need targeted interventions to help students prepare effectively. 

What stands out is the high adoption of Jill Watson in these courses, reinforcing the premise that students view Jill as a valuable tool for navigating course content. This success can be attributed to Jill's advanced design, which addresses some of the inherent limitations of LLMs. By using RAG to restrict context and incorporating advanced prompting, Jill ensures relevant and safe answers to student questions while suppressing biases and hallucinations.

\subsection{Complexity of Questions}


\begin{table}[]
\centering
\begin{tabular}{|l|l|l|l|}
\hline
\textbf{Cognitive Dimension} & \textbf{KBAI} & \textbf{CogSci} & \textbf{English} \\ \hline
Knowledge                    & 51.4\%         & 56.7\%           & 53.9\%            \\ \hline
Comprehension                & 9.3\%          & 6.2\%            & 10.7\%            \\ \hline
Application                  & 3.1\%          & 4.0\%            & 6.2\%             \\ \hline
Analysis                     & 27.8\%         & 15.7\%           & 10.7\%            \\ \hline
Evaluation                   & 4.6\%          & 8.7\%            & 9.9\%             \\ \hline
Synthesis                    & 3.9\%          & 8.5\%            & 8.6\%             \\ \hline
\end{tabular}
\caption{Cognitive Dimension of Synthetically Generated Questions for Spring 2024 (as percentages)}
\label{tab:spring2024_synth}
\end{table}
\begin{table}[]
\centering
\begin{tabular}{|l|l|l|l|}
\hline
\multicolumn{1}{|c|}{\textbf{Cognitive Dimension}} & \multicolumn{1}{c|}{\textbf{KBAI}} & \multicolumn{1}{c|}{\textbf{CogSci}} & \multicolumn{1}{c|}{\textbf{Eng}} \\ \hline
Knowledge                                          & 35.89\%                             & 61.45\%                               & 50.00\%                            \\ \hline
Comp                                               & 15.94\%                             & 15.96\%                               & 12.50\%                            \\ \hline
Application                                        & 7.24\%                              & 5.12\%                                & 3.12\%                             \\ \hline
Analysis                                           & 24.84\%                             & 10.54\%                               & 3.12\%                             \\ \hline
Evaluation                                         & 7.86\%                              & 3.31\%                                & 18.75\%                            \\ \hline
Synthesis                                          & 8.23\%                              & 3.61\%                                & 12.50\%                            \\ \hline
\end{tabular}
\caption{Cognitive Dimension of Real Student Questions for Spring 2024 (as percentages)}
\label{tab:percentage_real}
\end{table}

\begin{table}[]
\centering
\begin{tabular}{|l|l|l|l|l|l|}
\hline
\textbf{\begin{tabular}[c]{@{}l@{}}Cognitive \\ Dimension\end{tabular}}      & \textbf{\begin{tabular}[c]{@{}l@{}}Sp24\\ KBAI\end{tabular}} & \textbf{\begin{tabular}[c]{@{}l@{}}Sp24\\ CogSci\end{tabular}} & \textbf{\begin{tabular}[c]{@{}l@{}}Sp24\\ Eng\end{tabular}} & \textbf{\begin{tabular}[c]{@{}l@{}}F23\\ KBAI\end{tabular}} & \textbf{\begin{tabular}[c]{@{}l@{}}F23\\ Eng\end{tabular}} \\ \hline
Knowledge                                                                    & 689                                                          & 204                                                            & 16                                                          & 830                                                         & 105                                                        \\ \hline
Comp                                                                         & 306                                                          & 53                                                             & 4                                                           & 319                                                         & 9                                                          \\ \hline
Application                                                                  & 139                                                          & 17                                                             & 1                                                           & 341                                                         & 5                                                          \\ \hline
Analysis                                                                     & 477                                                          & 35                                                             & 1                                                           & 555                                                         & 6                                                          \\ \hline
Evaluation                                                                   & 151                                                          & 11                                                             & 6                                                           & 154                                                         & 27                                                         \\ \hline
Synthesis                                                                    & 158                                                          & 12                                                             & 4                                                           & 112                                                         & 11                                                         \\ \hline
\textbf{\begin{tabular}[c]{@{}l@{}}Total number\\ of questions\end{tabular}} & \textbf{1920}                                                & \textbf{332}                                                   & \textbf{32}                                                 & \textbf{2311}                                               & \textbf{163}                                               \\ \hline
\end{tabular}
\caption{Counts of total student questions and number of questions for each Cognitive Dimension for Spring 2024 (Sp24) and Fall 2023 (F23)}
\label{tab:dim_real}
\end{table}

First, we look at the complexity of the questions that Jill is tested on, pre-deployment, to set a baseline of what we expected Jill to be primarily used for. Next, to assess actual usage, we will look at the complexity of the real student questions collected.

We classified all questions in the CQAs using the fine-tuned BERT model described in the previous section. Since we explicitly prompt PaLM API to generate a variety of questions as part of the CQAs for testing, we expect the CQAs to somewhat reflect types of questions that would naturally arise from the courseware. 
Table \ref{tab:spring2024_synth} shows the distribution of these test questions across the six cognitive levels for each of the three courses Jill was deployed in Spring 2024.
We observe that a majority of the questions fall under the 'Knowledge' category or deal with recalling or recognizing factual information. This provides a baseline for the kind of questions Jill has tested on and the expectation that Jill Watson would perform well as an information retrieval tool.

\begin{table}[]
\centering
\begin{tabular}{|l|l|}
\hline
\textbf{Dimension} & \textbf{Real Student Question}                                                                                                                                \\ \hline
Knowledge          & \begin{tabular}[c]{@{}l@{}}Can you help me define heuristics and \\ give me an example?\end{tabular}                                                                                                                                                                                                                                                                                       \\ \hline
Comp               & \begin{tabular}[c]{@{}l@{}}Can you summarize the second chapter \\ for me?\end{tabular}                                                                                                                                                                                                                                                                                                    \\ \hline
Application        & \begin{tabular}[c]{@{}l@{}}Would rotating an image or finding \\ symmetry in an image be considered \\ conceptual or imagistic?\end{tabular}                                                                                                                                                                                                                                               \\ \hline
Analysis           & \begin{tabular}[c]{@{}l@{}}What is the difference between control \\ knowledge and heuristics?\end{tabular}                                                                                                                                                                                                                                                                                \\ \hline
Evaluation         & \begin{tabular}[c]{@{}l@{}}Fear can be interpreted as a shifting of \\ the allocation of computational resources\\  to expand CRUM. Computational \\ resources, such as memory, attention, etc.,\\ are limited and fear can allocate those \\ precious resources to the urgent situation \\ that is relevant to surviving goal. Is there \\ anything wrong in my proposition?\end{tabular} \\ \hline
Synthesis          & \begin{tabular}[c]{@{}l@{}}Consider an extended version of the \\ scientific method that incorporates \\ conceptual and simulation models. How \\ might scientists pairing conceptual models \\ with simulation models of the same \\ concepts be used as part of the \\ scientific method?\end{tabular}                                                                                   \\ \hline
\end{tabular}
\caption{Examples of real student questions per cognitive dimension, showing the evolving complexity of questions asked by students.}
\label{tab:examples_level}
\end{table}


Table \ref{tab:percentage_real} shows the percentage of real student questions within each cognitive level.
The numbers show that the ‘Knowledge’ dimension contains the highest portion of questions across all courses. This aligns with our expectation that Jill Watson could serve as an advanced information retrieval tool in classrooms. 
We also notice that, in general, there was a notable presence of higher-order cognitive questions, such as those requiring analysis and evaluation, challenging the tool to deliver beyond basic factual responses.
These numbers are observed to vary by course, leading us to speculate whether how a course is structured plays a role in this distribution.

To examine the relationship between course/semester and the cognitive dimensions of student questions, we ran a chi-squared test of independence (\(\chi^2(15) = 194.53\), \(p < 0.0001\)) on the question frequencies for each dimension, as seen in Table \ref{tab:dim_real}, suggesting that the distribution of question types across cognitive dimensions is indeed strongly dependent on the course and semester.
Table \ref{tab:examples_level} shows some of the real student questions per cognitive dimension to further shed a light on the type of questions that students have been asking Jill.

\begin{figure*}
\centering
\Description{Question distribution during exam weeks in KBAI showing that most critical questions occur during exam weeks.}
\includegraphics[width=2.75in]{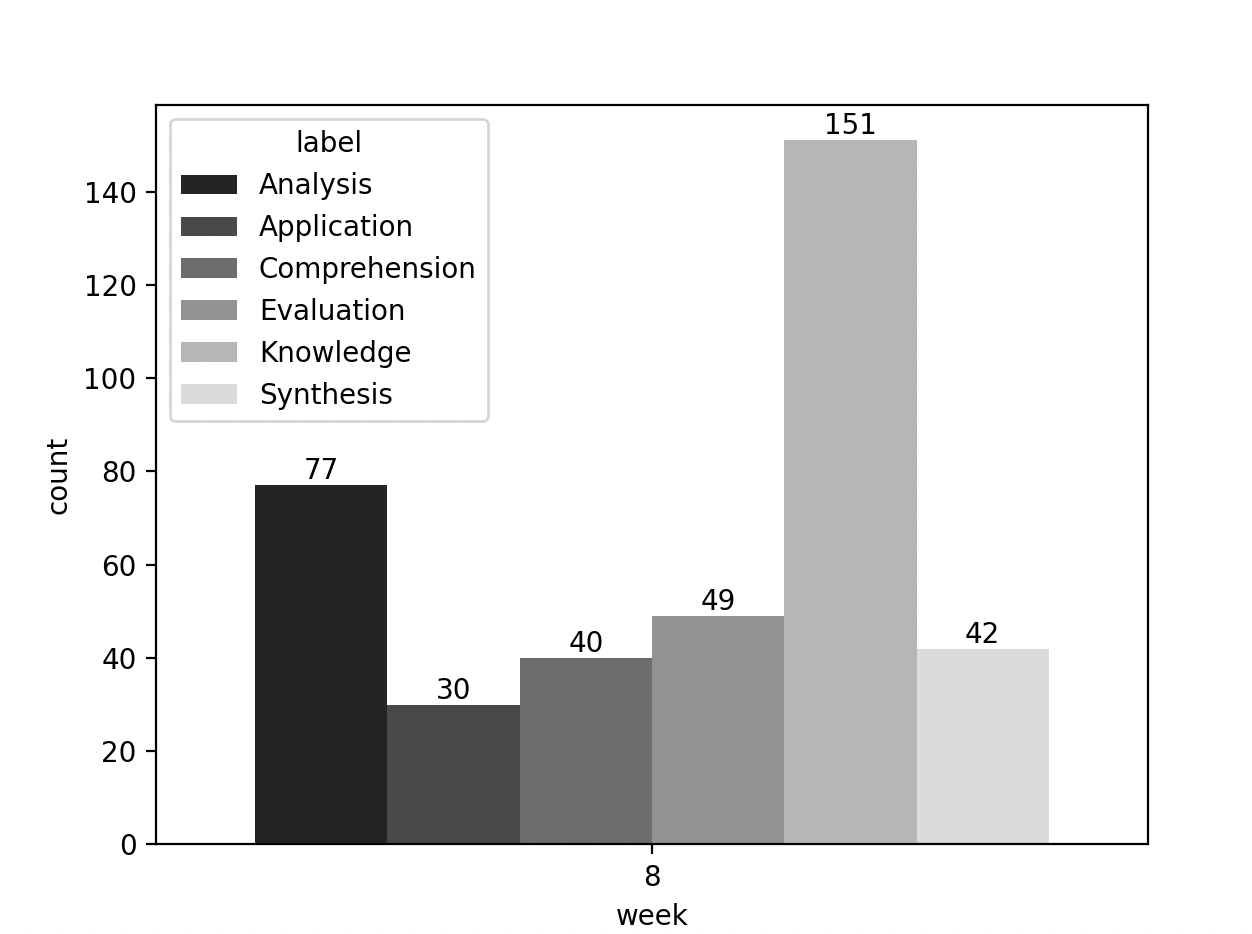}
\includegraphics[width=2.75in]{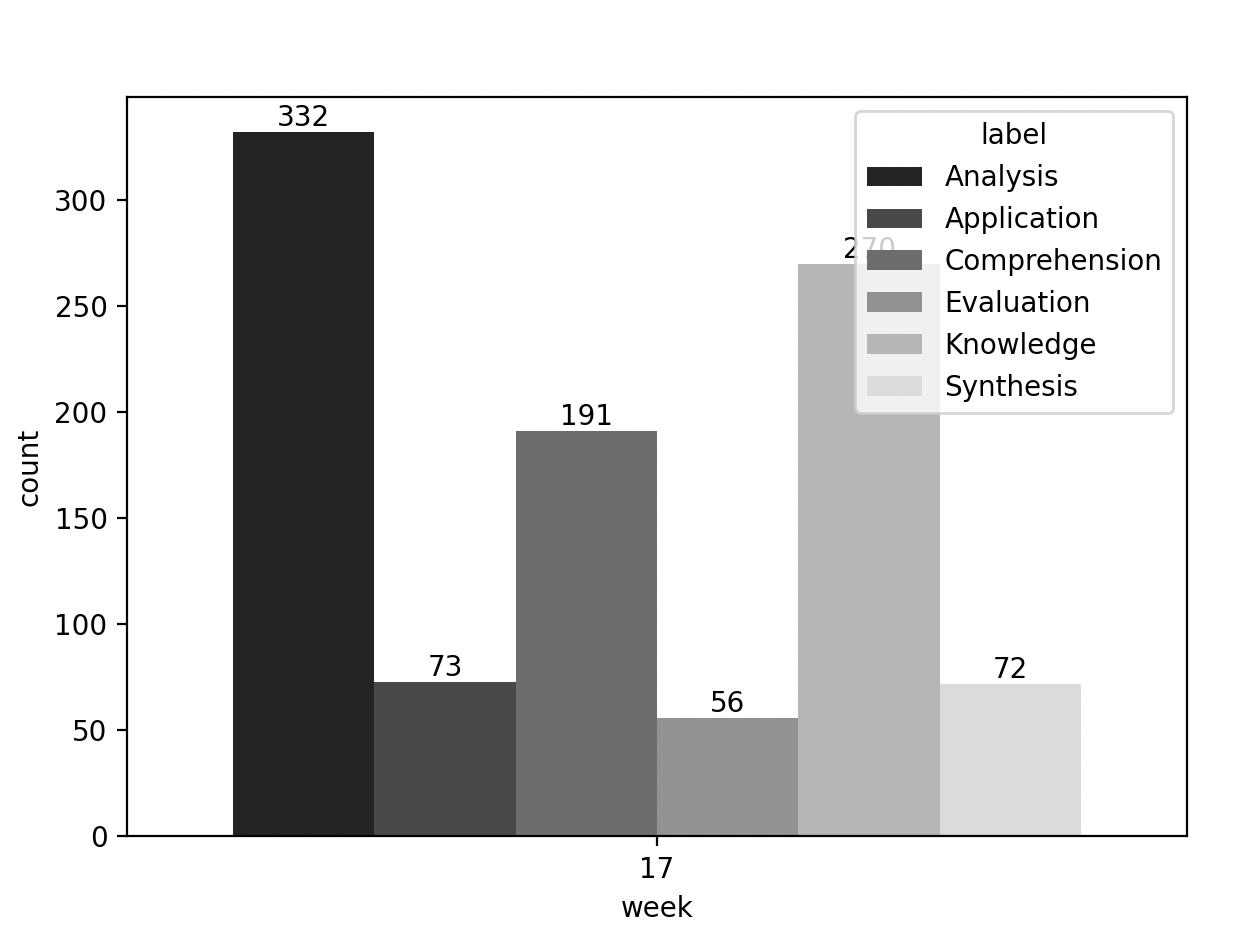}
\caption{Question distribution during exam weeks in KBAI: While there is a high number of factoid or 'Knowledge' type questions, numbers indicate that higher order or 'critical' questions occur more during high-pressure periods such as exams.}
\label{fig:exam weeks}
\end{figure*}
When analyzing the usage and complexity of questions over the entire semester, we also notice a relatively uniform distribution of questions in case of Intro to Cognitive Science, which is inline with how critical deliverables are distributed throughout the semester.
In contrast, we have previously seen that in case of KBAI, there is a spike in Jill Watson usage during the exam weeks. Figure \ref{fig:exam weeks} illustrates that this period also corresponds with a surge in critical questions from students.
This leads us to believe that the way a course is structured has an influence on the critical question-asking behavior of students.
However, further studies through student survey will be required to quantify the dependence between course structure and critical-question asking behavior. 

It is important to note that the number of questions asked in the English course is significantly lower than that in the KBAI or Intro to Cognitive Science course. Therefore, the observed distribution of questions and variations between semesters for this course may not necessarily indicate a substantive dependence on the pedagogical focus. 




\subsection{Qualitative Analysis of Responses}
The integration of Jill Watson with LLMs has introduced several desirable characteristics in its responses which could further have an effect on usage. Qualitative analysis of the responses generated by Jill Watson reveals structured and educational patterns in its output such as:

\begin{table}[]
\centering
\begin{tabular}{|l|}
\hline
\begin{tabular}[c]{@{}l@{}}Student1: What is universal AI vs specialized AI\\ \\ Jill: Universal AI and specialized AI represent two \\ different approaches to artificial intelligence:\\ 1. Universal AI: Universal AI aims to create intelligent \\ systems that can perform a wide range of tasks and adapt \\ to various domains without significant reprogramming or\\ retraining. [...]\\ 2. Specialized AI: Specialized AI, on the other hand, \\ focuses on developing AI systems that excel at specific \\ tasks or functions within a narrow domain. These systems \\ are designed to be highly optimized for particular \\ applications, such as image recognition, natural language \\ processing, or autonomous driving. [...]\\ \\ In summary, universal AI aims for broad adaptability and\\ general intelligence across diverse tasks, while \\ specialized AI targets high performance and efficiency \\ within specific domains or applications. The choice \\ between universal and specialized AI depends on the \\ desired scope of functionality and the specific \\ requirements of the problem or task at hand.\end{tabular} \\ \hline
\end{tabular}
\caption{Excerpt from a real student interaction with Jill Watson in the KBAI course: Jill's response includes definitions, examples, and a summary to reinforce understanding.}
\label{tab:conv1}
\end{table}  

\begin{itemize}
    \item Definition and Examples: Responses start with a clear definition of the concept, followed by relevant examples, reinforcing theoretical points with practical instances.
    \item Comparison and Contrast: Another frequent response pattern involves comparing and contrasting different concepts, which is systematically followed by examples. This method aids students in understanding differences and similarities between concepts, fostering deeper analytical skills.
    \item Summary of Key Points: Responses typically conclude with a summary of key points, ensuring clarity and retention of the discussed information.
    \item Adaptive Tone and Sentiment: Responses are tailored to the tone and sentiment of the student's query, maintaining politeness and engagement, which is specifically prompted to the agent.
\end{itemize}      

Tables \ref{tab:conv1} show one such real student interaction with Jill Watson that supports our observations listed above.   

\subsection{Student Feedback to Jill's Responses}
The Jill Watson interface allows students to provide feedback for Jill's responses by marking them as either helpful or not helpful. Across all three courses, we observed a substantial amount of positive feedback, as seen in Table \ref{tab:feedback}. Specifically, in the KBAI course, 191 responses (78\% of responses that received feedback) were marked helpful compared to 53 marked not helpful. In the Cognitive Science course, 47 responses (67\% of responses that received feedback) were marked as helpful while 24 were not, while for the English course, 17 (94\% of responses that received feedback) were marked helpful and 1 marked not helpful. Responses that are marked as helpful typically contain elements such as examples and summaries, which aid in understanding the material. The consistently high positive feedback across different courses highlights the effectiveness of Jill Watson in providing valuable assistance to students.

\begin{table}[]
\centering
\begin{tabular}{|l|l|l|}
\hline
\multicolumn{1}{|c|}{\textbf{Course}} & \multicolumn{1}{c|}{\textbf{Positive feedback}} & \multicolumn{1}{c|}{\textbf{Negative feedback}} \\ \hline
KBAI                                  & 191                                                  & 53                                                   \\ \hline
CogSci                                & 47                                                   & 24                                                   \\ \hline
English                               & 17                                                   & 1                                                    \\ \hline
\end{tabular}
\caption{Number of Jill's responses marked as helpful (Positive) and not helpful (Negative) by students in Spring 2024.}
\label{tab:feedback}
\end{table}

\section{Limitations}
Our study has several limitations that must be acknowledged and that pave the way for future work. While we have identified a trend towards higher-order question-asking in certain course structures, the study does not deeply explore the underlying factors driving these behaviors. More granular investigations into how specific pedagogical approaches, course designs and external factors like digital literacy influence student interaction with AI tools are necessary to draw more precise conclusions.
Second, while our analysis covers a range of question types and cognitive levels, it does not fully address the quality of the responses provided by Jill Watson. Future studies should include a more detailed assessment of response accuracy and pedagogical soundness to ensure that the tool not only engages students but also supports effective learning.



\section{Conclusions}
Our study shows that incorporating LLMs into Jill Watson has led to increased tool usage and exploration among students under varying settings. We have observed that a significant portion of student inquiries are related to course content, indicating that students are utilizing Jill as a resource for understanding classroom material. We also see a notable presence of questions with high cognitive complexity, suggesting that students enagaged deeply with the material and used Jill for more than just fact-seeking. Although further work is needed, our qualitative analysis of Jill's responses highlights several ideal features that may encourage students to ask more critical questions. Jill's capability to address questions of varying complexities, ensure the relevance and safety of responses, and mitigate biases and hallucinations through the implementation of RAG demonstrates the potential of AI-based educational tools to engage students in meaningful and deeper learning experiences.


The frequency and complexity of these higher-order questions vary significantly with the course structure, suggesting that the way a course is designed and implemented can influence how students utilize AI-based educational tools. Courses that emphasize problem-solving tend to elicit a higher frequency of higher-order cognitive questions, whereas courses focused on writing primarily generate questions centered on understanding and comprehension. Moreover, the utility of Jill evolves over the semester, indicating that adapting to student needs is a critical factor in design of such tools.

\section{Acknowledgments}

This research has been supported by NSF Grants \#2112532 and \#2247790 to the National AI Institute for Adult Learning and Online Education. This study has been conducted in accordance with approved IRB protocols. We thank the members of the Design Intelligence Laboratory for their contributions to this work. We are especially grateful to Sandeep Kakar of the Jill Watson team.

\bibliographystyle{abbrv}
\bibliography{vtagpt}  
%

\balancecolumns
\end{document}